\documentclass[pre,aps,preprint,showpacs,preprintnumbers,amsmath,amssymb,secnumroman,eqsecnum,eqappnum]{revtex4}

\usepackage{graphicx}% Include figure files
\usepackage{dcolumn}% Align table columns on decimal point
\usepackage{bm}% bold math

\newcommand{\beq}{\begin{equation}}
\newcommand{\eeq}{\end{equation}}
\newcommand{\beqa}{\begin{eqnarray}}
\newcommand{\eeqa}{\end{eqnarray}}
\newcommand{\vc}[1]{\mbox{\boldmath $#1$}}

\newcommand{\vol}[1]{{\bf #1}}

\begin{document}

%\preprint{APS/123-QED}

\title{Dynamics of cruising swimming by a deformable sphere for two simple models}% Force line breaks with \\

\author{B. U. Felderhof}
 %\altaffiliation[Also at ]{Physics Department, XYZ University.}%Lines break automatically or can be forced with \\

 \email{ufelder@physik.rwth-aachen.de}
\affiliation{Institut f\"ur Theorie der Statistischen Physik\\ RWTH Aachen University\\
Templergraben 55\\52056 Aachen\\ Germany\\
}%

\author{R. B. Jones}
 %\altaffiliation[Also at ]{Physics Department, XYZ University.}%Lines break automatically or can be forced with \\

 \email{r.b.jones@qmul.ac.uk}
\affiliation{Queen Mary University of London, The School of
Physics and Astronomy, Mile End Road, London E1 4NS, UK\\}%

\date{\today}% It is always \today, today,
             %  but any date may be explicitly specified

\begin{abstract}
The dynamics of periodic swimming is studied for two models of a deformable sphere, the dipole-quadrupole model and the quadrupole-octupole model. For the two models the solution of the Navier-Stokes equations can be found exactly to second order in the amplitude of stroke. Hence the oscillating force exerted by the fluid on the body is calculated. This allows calculation of the periodic time-dependent center of mass velocity.
\end{abstract}

% PACS, the Physics and Astronomy
                             % Classification Scheme.
%\keywords{Suggested keywords}%Use showkeys class option if keyword
                              %display desired
\maketitle
\section{\label{I}Introduction}

The swimming of fish and the flying of birds pose a challenging problem of fluid mechanics \cite{1}-\cite{8}. Many other organisms in nature on a wide range of scales perform locomotion in a fluid environment. In the present article we restrict ourselves to a study of rectilinear cruising motion with a repetitive periodic flow pattern in infinite viscous incompressible fluid with flow described by the Navier-Stokes equations. We assume a stroke of small amplitude and neglect wake formation and turbulence.

Even with these simplifying assumptions the problem is difficult. In theory it requires the solution of the nonlinear Navier-Stokes equations with no-slip boundary condition imposed on a body with periodically changing shape. For given shape deformations one would like to determine the time-dependent swimming velocity and the work required to perform the motion.

A popular approach is based on the thrust-drag paradigm \cite{1}. In this approach the force exerted on the fluid is calculated from the surface deformations of the body and subsequently balanced by the frictional drag exerted by the fluid on the body. The swimming velocity is evaluated from the balance equation. Simple mechanical models show that this approach is inappropriate, since both the mean force and the mean drag vanish when averaged over a period \cite{9}-\cite{12}. The models consist of structures of spheres with elastic and hydrodynamic interactions, as given by a friction matrix and an added mass matrix. The structures change in form due to actuating forces acting on the spheres. The actuating forces oscillate in time and add up to zero at any time. Nonetheless by a judicious combination of forces the structures perform locomotion.

The mechanical models provide valuable insight, but do not do justice to the full complexity of hydrodynamic interactions. The Navier-Stokes equations take complete account of these interactions. The model of a deformable sphere immersed in fluid is sufficiently simple that at least some analytical progress can be made. The model was first studied by Lighthill \cite{13} and Blake \cite{14} in the Stokes limit of high friction. Later it was extended to incorporate the effect of fluid inertia \cite{15}-\cite{18}. The relative importance of friction and fluid inertia is measured by the dimensionless scale number $s$ defined by $s^2=a^2\omega\rho/(2\eta)$, where $a$ is the sphere radius, $\omega$ is the frequency, $\rho$ is the mass density of the fluid, and $\eta$ is the shear viscosity. The Stokes limit corresponds to $s=0$, and the limit $s\rightarrow\infty$ is dominated by inertia. Recently we devised two simple models \cite{19}, the dipole-quadrupole model (DQ) and the quadrupole-octupole model (QO) for which the fluid motion can be studied in detail to second order in the amplitude of stroke over the full range of scale number $s$.

In the DQ and QO models the flow is chosen to be irrotational to first order in the amplitude of stroke. As a consequence, in second order the Reynolds stress is balanced by a pressure of Bernoulli type and need not be further considered. This property leads to a significant simplification of the analysis. The models are studied not as representation of actual organisms. Rather, they are chosen as paradigmatic examples for which questions of principle can be elucidated.

In the two models the inner structure of the body is not considered. Its shape deformations are prescribed kinematically. As a consequence the time-dependent swimming velocity pertains to the net motion of the body surface. The motion of the body center of mass can in principle be determined via Newton's equation from the time-dependent force exerted by the fluid on the body. It is therefore of interest to study the force corresponding to the prescribed surface distortions. At the same time the investigation throws light on the thrust-drag paradigm.

It turns out that in the DQ and QO models the calculation of the pressure and viscous forces is quite subtle. The final expression for the total time-dependent force depends in an interesting way on viscosity and fluid mass density. The total force to second order in amplitude oscillates in time at twice the frequency and vanishes on time average. We discuss also the work, dissipation, and the efficiency of swimming for the two models. In Sec. V we discuss the relevance of the internal dynamic structure of the body. It turns out that for the DQ model one must require that the first order oscillating force on the body is fully absorbed internally, and has negligible effect on the motion of the surface. In Sec. VI we discuss the thrust-drag paradigm.

\section{\label{II}Forces exerted on the fluid and on the body}

We consider a swimmer immersed in a viscous
incompressible fluid of shear viscosity $\eta$ and mass density $\rho$.
The fluid is assumed to be infinite in all directions. It is set in motion by time-dependent distortions of the swimmer. Its body is assumed to be simply-connected with an axisymmetric shape $S(t)$ which varies periodically in time with period $T=2\pi/\omega$ such that it leads to swimming in the direction of the axis, which is identified with the $z$ axis of a Cartesian sytem of coordinates. At the surface $S(t)$ the fluid flow velocity satisfies the no-slip boundary condition. The flow velocity $\vc{u}(\vc{r},t)$ and pressure $p(\vc{r},t)$ are assumed to satisfy the Navier-Stokes equations
\begin{equation}
\label{2.1}\rho\bigg[\frac{\partial\vc{u}}{\partial t}+\vc{u}\cdot\nabla\vc{u}\bigg]=\eta\nabla^2\vc{u}-\nabla p,\qquad\nabla\cdot\vc{u}=0.
\end{equation}
We consider a solution of these equations which varies periodically in time, as caused by the periodically varying shape of the body. In the laboratory frame the flow velocity tends to zero at infinity and the pressure tends to the ambient value $p_0$. The periodicity of the solution implies
\begin{eqnarray}
\label{2.2}\vc{u}(\vc{r}+\overline{U}T\vc{e}_z,t+T)&=&\vc{u}(\vc{r},t),\nonumber\\
p(\vc{r}+\overline{U}T\vc{e}_z,t+T)&=&p(\vc{r},t),
\end{eqnarray}
where $\overline{U}$ is the mean swimming velocity, and $\vc{e}_z$ is the unit vector in the $z$ direction.
The flow $(\vc{u}(\vc{r},t)-\overline{U}\vc{e}_z,p(\vc{r},t))$ in the rest frame can be decomposed into Fourier components oscillating at frequencies $n\omega$, where $n$ runs through all integers. The steady state component at zero frequency is denoted as $(\vc{u}'(\vc{r}),\;p'(\vc{r}))$. This is called the net flow.

The flow velocity at large distance from the body can be expanded into vector spherical harmonics and the pressure disturbance can be expanded into scalar spherical harmonics centered at the centroid of the body at time $t$. We assume that the shape changes are such that the expansions contain no monopole component. A monopole component would correspond to a spherical expansion of volume. For a flow with a dipole component the total fluid momentum $\vc{P}_f$, given by the integral of $\rho\vc{u}$ over the fluid, is only conditionally convergent. However, the rate of change of the total fluid momentum is well-defined. It is given by
\begin{equation}
\label{2.3}\frac{d\vc{P}_f}{dt}=-\vc{K}-\vc{\mathcal{L}},
\end{equation}
where $\vc{K}$ is the force exerted by the fluid on the body, and $\vc{\mathcal{L}}$ is the loss at infinity due to momentum flow.  The force is given by \cite{20}
\begin{equation}
\label{2.4}\vc{K}(t)=-\int_{S(t)}\vc{\Pi}\cdot\vc{n}\;dS,
\end{equation}
where $\vc{\Pi}$ is the momentum flow tensor and $\vc{n}$ is the normal to the body pointing into the fluid. The tensor $\vc{\Pi}$ is given by
\begin{equation}
\label{2.5}\vc{\Pi}=\rho\vc{u}\vc{u}+p\vc{I}-\vc{\sigma}_v,
\end{equation}
with unit tensor $\vc{I}$ and viscous stress tensor
\begin{equation}
\label{2.6}\vc{\sigma}_v=\eta[\nabla\vc{u}+(\nabla\vc{u})^T)].
\end{equation}
The Navier-Stokes equations Eq. (2.1) can be written alternatively as
\begin{equation}
\label{2.7}\rho\frac{\partial\vc{u}}{\partial t}+\nabla\cdot\vc{\Pi}=0,\qquad\nabla\cdot\vc{u}=0.
\end{equation}
The loss term in Eq. (2.3) is given by
\begin{equation}
\label{2.8}\vc{\mathcal{L}}(t)=\lim_{R\rightarrow\infty}\int_{S_R}\vc{\Pi}\cdot\vc{n}_R\;dS_R,
\end{equation}
where the integration is over a sphere of radius $R$ centered at the centroid of $S(t)$ and $\vc{n}_R$ is the outward normal to the sphere. The tensor $\vc{\Pi}$ tends to $p_0\vc{I}$ at infinity, where $p_0$ is the constant ambient pressure. The corresponding contribution to the integral in Eq. (2.8) vanishes.
The three vectors $\vc{P}_f$,$\vc{K}$, and $\vc{\mathcal{L}}$  point in the $z$ direction and vary periodically in time.

Averaging Eq. (2.3) over a period we find
\begin{equation}
\label{2.9}\overline{\vc{K}}=\frac{1}{T}\int^T_0\vc{K}(t)\;dt=-\overline{\vc{\mathcal{L}}},
\end{equation}
since for periodic motion $\vc{P}_f(T)=\vc{P}_f(0)$.
We define separately
\begin{equation}
\label{2.10}\vc{K}_R(t)=-\int_{S(t)}\rho\vc{u}\vc{u}\cdot\vc{n}\;dS,\qquad\vc{K}_p(t)=-\int_{S(t)}p\vc{n}\;dS,\qquad\vc{K}_v(t)=\int_{S(t)}\vc{\sigma}_v\cdot\vc{n}\;dS.
\end{equation}
We call the first term the Reynolds force, the second the pressure force, and the third the viscous force.

In earlier work \cite{19} we considered two simple models of a sphere of radius $a$ swimming in a viscous incompressible fluid by surface distortion, the dipole-quadrupole and the quadrupole-octupole swimmer. We evaluated the swimming velocity and the flow pattern of both swimmers to second order in the amplitude of the stroke \cite{19}. More generally we consider swimmers with stroke given by the surface distortion
\begin{equation}
\label{2.11}\vc{\xi}(\vc{s},t)=\varepsilon a\big[\mu_l\vc{B}_l(\theta)\sin(\omega t)-\mu_{l+1}\;\vc{B}_{l+1}(\theta)\cos(\omega t)\big],
\end{equation}
where $\vc{s}=a\hat{\vc{r}}$ denotes a point on the undistorted sphere corresponding to polar angle $\theta$, and $\vc{B}_l$ are vector spherical harmonics defined by
 \begin{equation}
\label{2.12}
\vc{B}_l=-(l+1)P_l\vc{e}_r+\frac{\partial P_l}{\partial\theta}\;\vc{e}_\theta,
\end{equation}
with unit vectors $\vc{e}_r=\hat{\vc{r}},
\;\vc{e}_\theta$, and Legendre polynomials $P_l(\cos\theta)$ in the notation of Edmonds \cite{21}. On the righthand side of Eq. (2.11) $\varepsilon$ is an amplitude factor, and $\mu_l$ is a dimensionless coefficient. The first order flow corresponding to Eq. (2.11) is
\begin{eqnarray}
\label{2.13}\vc{v}^{(1)}(\vc{r},t)&=&\varepsilon a\omega\bigg[\mu_l\frac{a^{l+2}}{r^{l+2}}\;\vc{B}_l(\theta)\cos(\omega t)+\mu_{l+1}\frac{a^{l+3}}{r^{l+3}}\;\vc{B}_{l+1}(\theta)\sin(\omega t)\bigg],\nonumber\\
p^{(1)}(\vc{r},t)&=&\varepsilon\rho a^2\omega^2\bigg[\mu_l\frac{a^{l+1}}{r^{l+1}}\;P_l(\cos\theta)\sin(\omega t)-\mu_{l+1}\frac{a^{l+2}}{r^{l+2}}\;P_{l+1}(\cos\theta)\cos(\omega t)\bigg].
\end{eqnarray}
For the dipole-quadrupole swimmer $l=1$, and for the quadrupole-octupole swimmer $l=2$. The coordinates $(r,\theta,\varphi)$ are defined with the center of the sphere at the origin. The first expression in Eq. (2.13) implies that the first order velocity of the sphere vanishes, $\vc{U}^{(1)}=0$.

The dipolar contribution to the pressure decays as $1/r^2$. The momentum of the dipolar part of the flow $\vc{v}^{(1)}$ in the volume $a<r<R$ vanishes by angular integration. To first order the force and the momentum loss are given by
\begin{equation}
\label{2.14}\vc{K}^{(1)}(t)=-\vc{\mathcal{L}}^{(1)}(t)=-\frac{4\pi}{3}\varepsilon\mu_1a^4\rho\omega^2\sin(\omega t)\vc{e}_z.
\end{equation}
This does not contribute to the mean force $\overline{\vc{K}}$ or mean momentum loss $\overline{\vc{\mathcal{L}}}$. 

Since the first order velocity field $\vc{v}^{(1)}(\vc{r},t)$ is both irrotational and incompressible, the first order viscous stress tensor $\vc{\sigma}^{(1)}_v$ has vanishing divergence, $\nabla \cdot \vc{\sigma}^{(1)}_v =0$. By using Gauss's theorem to convert the expression for $\vc{K}_v(t)$ in Eq. (2.10) into an integral of  $\nabla \cdot \vc{\sigma}^{(1)}_v$ over the fluid volume one sees that there is no force contribution from the viscous part of the stress tensor $\vc{\sigma}^{(1)}_v$.

From Eq. (2.1) we see that the second order  Navier-Stokes equations for the fields $(\vc{v}^{(2)}, p^{(2)})$ have an inhomogeneous term invoving the first order irrotational field   $\vc{v}^{(1)}$  in the form of the Reynolds force density $\rho \vc{v}^{(1)}  \cdot \nabla \vc{v}^{(1)} $. However, remembering that $\vc{v}^{(1)}$  is a potential flow and by expressing the pressure as $ p^{(2)} =  p^{(2)} _S+ p_B$ with the Bernoulli type pressure $p_B = -\frac{1}{2} \rho {\vc{v}^{(1) 2}}$, the Reynolds force density is exactly cancelled by $p_B$ \cite{15}  leaving a homogeneous linear equation for the second order fields $(\vc{v}^{(2)}, p^{(2)}_S)$ whose solution can be found in full detail in Felderhof and Jones\cite{19}. Thus there is no second order contribution to the net time-dependent forces from $\vc{K}_R (t)$ or $p_B$.

The remaining second order pressure and viscous contributions $\vc{K}^{(2)}_p$ and $\vc{K}^{(2)}_v$ to the time-dependent force can each be written as a sum of three terms
\begin{equation}
\label{2.15}\vc{K}^{(2)}_p=\vc{K}_{p20}+\vc{K}_{p11}+\vc{K}_{pd},
\qquad\vc{K}^{(2)}_v=\vc{K}_{v20}+\vc{K}_{v11}+\vc{K}_{vd},
\end{equation}
with separate contributions
\begin{eqnarray}
\label{2.16} \vc{K}_{p20}&=&-\int_{S_0}p^{(2)}_S\vc{n}_0\;dS_0,\qquad\vc{K}_{v20}=\int_{S_0}\vc{\sigma}^{(2)}_v\cdot\vc{n}_0\;dS_0,\nonumber\\
\vc{K}_{p11}&=&-\int_{S_0}p^{(1)}(\vc{n}\;dS)^{(1)},\qquad\vc{K}_{v11}=\int_{S_0}\vc{\sigma}^{(1)}_v\cdot(\vc{n}\;dS)^{(1)},\nonumber\\
\vc{K}_{pd}&=&-\int_{S_0}(\vc{\xi}\cdot\nabla p^{(1)})\vc{n}_0\;dS_0,\qquad\vc{K}_{vd}=\int_{S_0}(\vc{\xi}\cdot\nabla\vc{\sigma}^{(1)}_v)\cdot\vc{n}_0\;dS_0,
\end{eqnarray}
where in the first line $p^{(2)}_S$ is the additional second order pressure due directly to surface modulations, and in the second line we integrate over the first order correction to  the surface element $\vc{n}_0dS_0$. From differential geometry \cite{22},\cite{23} one finds that for a sphere $S_0$ with axisymmetric distortion $(\xi_r,\xi_\theta,0)$ the first order surface element is
\begin{equation}
\label{2.17}(\vc{n}\;dS)^{(1)}=a\bigg(2\xi_r\sin\theta+\xi_\theta\cos\theta+\frac{\partial\xi_\theta}{\partial\theta}\sin\theta,
\big(\xi_\theta-\frac{\partial\xi_r}{\partial\theta}\big)\sin\theta,0\bigg)d\theta d\varphi
\end{equation}
in spherical coordinates $(r,\theta,\varphi)$. The third line in Eq. (2.16) takes account of the displacement of the surface of integration in Eq. (2.4). In the next section we consider the various contributions in Eq. (2.16).

\section{\label{III}Force calculations}

For the two models under consideration the various contributions in Eq. (2.16) can be evaluated in explicit form. By symmetry all forces point in the $z$ direction. The time average of the first two forces in Eq. (2.16) vanishes,
\begin{equation}
\label{3.1} \overline{\vc{K}_{p20}}=-\int_{S_0}\overline{p^{(2)}_S}\vc{n}_0\;dS_0=0,
\qquad\overline{\vc{K}_{v20}}=\int_{S_0}\overline{\vc{\sigma}^{(2)}_v}\cdot\vc{n}_0\;dS_0=0.
\end{equation}
For the last two contributions to the pressure force we find
\begin{eqnarray}
\label{3.2}K_{p11}(t)=-\frac{12}{5}\pi\varepsilon^2\mu_1\mu_2a^4\rho\omega^2\sin(2\omega t),\qquad (DQ)\nonumber\\
K_{p11}(t)=-\frac{24}{7}\pi\varepsilon^2\mu_2\mu_3a^4\rho\omega^2\sin(2\omega t),\qquad (QO)\nonumber\\
K_{pd}(t)=\frac{24}{5}\pi\varepsilon^2\mu_1\mu_2a^4\rho\omega^2\sin(2\omega t),\qquad (DQ)\nonumber\\
K_{pd}(t)=\frac{48}{7}\pi\varepsilon^2\mu_2\mu_3a^4\rho\omega^2\sin(2\omega t).\qquad (QO)\nonumber\\
\end{eqnarray}
Each of these forces vanishes on time average.

For the last two contributions to the viscous force we find
\begin{eqnarray}
\label{3.3}K_{v11}(t)=192\pi\varepsilon^2\mu_1\mu_2\eta a^2\omega\cos^2(\omega t),\qquad (DQ)\nonumber\\
K_{v11}(t)=480\pi\varepsilon^2\mu_2\mu_3\eta a^2\omega\cos^2(\omega t),\qquad (QO)\nonumber\\
K_{vd}(t)=-192\pi\varepsilon^2\mu_1\mu_2\eta a^2\omega\cos^2(\omega t),\qquad (DQ)\nonumber\\
K_{vd}(t)=-480\pi\varepsilon^2\mu_2\mu_3\eta a^2\omega\cos^2(\omega t).\qquad (QO)
\end{eqnarray}
This shows that in both cases we have the precise cancelation
\begin{equation}
\label{3.4}K_{v11}(t)+K_{vd}(t)=0.
\end{equation}
The forces do not vanish on time average, but for the sum one has $\overline{K_{v11}}+\overline{K_{vd}}=0$.

Adding the various contributions we find that the mean thrust to second order in amplitude $\overline{\vc{K}^{(2)}}$ vanishes. From Eq. (2.9) we find that the net loss also vanishes,
 \begin{equation}
\label{3.5}\overline{\vc{K}^{(2)}}=0,\qquad\overline{\vc{\mathcal{L}}^{(2)}}=0.
\end{equation}
In our earlier work \cite{15} we evaluated the mean swimming velocity from the condition that the swimmer exert no mean force on the fluid. The present work proves that the calculation \cite{19} is self-consistent.

Finally we consider the oscillating contributions to $\vc{K}_{p20}(t)$ and $\vc{K}_{v20}(t)$. For the pressure force we find
\begin{eqnarray}
\label{3.6}K_{p20}(t)&=&\frac{8\pi}{3}\varepsilon^2\mu_1\mu_2\eta a^3\omega\;\mathrm{Re}\bigg[\beta^2a\bigg(\frac{12}{5}+\frac{3k_2(\beta a)}{4k_0(\beta a)}\bigg)e^{-2i\omega t}\bigg],\qquad (DQ)\nonumber\\
K_{p20}(t)&=&\frac{8\pi}{3}\varepsilon^2\mu_2\mu_3\eta a^3\omega\;\mathrm{Re}\bigg[\beta^2a\bigg(\frac{30}{7}+\frac{3k_2(\beta a)}{2k_0(\beta a)}\bigg)e^{-2i\omega t}\bigg],\qquad(QO)
\end{eqnarray}
with modified spherical Bessel functions $k_l(z)$ and $\beta=(1-i)(\omega\rho/\eta)^{1/2}$. These forces are similar in nature to those given by Eq. (2.14).
For the viscous force we find
\begin{eqnarray}
\label{3.7}K_{v20}(t)&=&12\pi\varepsilon^2\mu_1\mu_2\eta a^2\omega\;\mathrm{Re}\big[(1+\beta a)e^{-2i\omega t}\big],\qquad (DQ)\nonumber\\
K_{v20}(t)&=&24\pi\varepsilon^2\mu_2\mu_3\eta a^2\omega\;\mathrm{Re}\big[(1+\beta a)e^{-2i\omega t}\big].\qquad(QO)
\end{eqnarray}
The factor $1+\beta a$ is familiar from the calculation of the friction coefficient of an oscillating sphere, where the same flow pattern occurs \cite{24}. The total second order time-dependent force acting on the body is
 \begin{equation}
\label{3.8}\vc{K}^{(2)}(t)=[K_{p11}(t)+K_{pd}(t)+K_{p20}(t)+K_{v20}(t)]\vc{e}_z.
\end{equation}
The oscillating pressure force $K_{p20}(t)$ is of particular interest. It corresponds to a long range pressure wave, as found from Eq. (3.1) of Ref. 19 with $l=1$,  falling off only as $1/r^2$ with distance, and leading to oscillating momentum loss $\vc{\mathcal{L}}^{(2)}(t)=-K_{p20}(t)\vc{e}_z$. Explicitly we find for the sum in Eq. (3.8)
\begin{eqnarray}
\label{3.9}K^{(2)}(t)&=&18\pi\varepsilon^2\mu_1\mu_2a^2\omega\bigg[\big(\eta+a\sqrt{\eta\rho\omega}\big)\cos(2\omega t)
-\big(a\sqrt{\eta\rho\omega}+\frac{4}{5}\;a^2\rho\omega\big)\sin(2\omega t)\bigg],\qquad (DQ)\nonumber\\
K^{(2)}(t)&=&36\pi\varepsilon^2\mu_2\mu_3a^2\omega\bigg[\big(\eta+a\sqrt{\eta\rho\omega}\big)\cos(2\omega t)
-\big(a\sqrt{\eta\rho\omega}+\frac{16}{21}\;a^2\rho\omega\big)\sin(2\omega t)\bigg].\qquad (QO)\nonumber\\
\end{eqnarray}
These expressions show an interesting dependence on viscosity and mass density.  One way to see this is to express Eq. (3.9) in terms of amplitude and phase as
\begin{equation}
\label{3.10}K^{(2)}(t) = P \cos (2 \omega t + \phi),
\end{equation}

\begin{eqnarray}
\label{3.11}
P&=& 18 \pi \epsilon^2 \mu_1 \mu_2 \eta a^2 \omega \left [ 1 + 2 \sqrt{2} s + 4 s^2 + \sqrt{2} \frac{16}{5} s^3 + \frac{64}{25} s^4 \right]^{\frac{1}{2}} ,\nonumber \\
 \tan \phi &=&\frac{\sqrt 2 s + \frac{8}{5} s^2} {1 + \sqrt 2 s} ,\qquad \qquad \qquad \qquad \qquad \qquad (DQ)  \nonumber \\
P&=& 36 \pi \epsilon^2 \mu_2 \mu_3 \eta a^2 \omega \left [ 1 + 2 \sqrt{2} s + 4 s^2 + \sqrt{2} \frac{64}{21} s^3 + \frac{1024}{441} s^4 \right]^{\frac{1}{2}} , \nonumber \\
  \tan \phi &=&\frac{\sqrt 2 s + \frac{32}{21} s^2} {1 + \sqrt 2 s} , \qquad \qquad \qquad \qquad \qquad \qquad(QO)
\end{eqnarray}
with  $s^2 = a^2 \omega \rho/(2 \eta)$. For small values of s (large viscosity or small mass density) we are in the Stokes flow limit and the phase angle is nearly zero while for large values of s (small viscosity or large mass density) the phase angle is nearly $\pi$/2. Thus in the Stokes limit the net force oscillates  as a second harmonic in phase with the second term in the surface displacement (2.11) while in the inertial limit of large s it oscillates as a second harmonic now in phase with the first term in the surface displacement. This is interesting because the second term of Eq. (2.11) involves more surface oscillation (as a function of $\theta$) than does the first term so that in a viscosity dominated situation the force is sensitive to the more rapid of the surface oscillations in $\theta$ while in an inertially dominated regime the force is more sensitive to the least deforming surface displacement.

A different aspect is revealed by studying separately the components $K_p^{(2)}$ and $K_v^{(2)}$ of the total force in Eq. (3.8) To illustrate this behavior numerically we define the force scale as
 \begin{equation}
\label{3.12}K_{0j}=6\pi\varepsilon^2\eta a^2\omega(1+s^2)\mu_j\mu_{j+1},\qquad (j=1,2).
\end{equation}
In Fig. 1 we plot the ratios $K^{(2)}_p(t)/K_{01}$ and $K^{(2)}_v(t)/K_{01}$, as well as their sum during one period for $s=0.01,\;s=1$, and $s=100$ for the DQ model. In Fig. 2 we plot the ratios $K^{(2)}_p(t)/K_{02}$ and $K^{(2)}_v(t)/K_{02}$, as well as their sum during one period for $s=0.01,\;s=1$, and $s=100$ for the QO model. The plots show that for both models for small and intermediate values of the scale number the pressure and viscous contributions to the second order force are of comparable magnitude, whereas for large scale number the pressure force dominates.

\section{\label{IV}Kinetic energy and power}

The kinetic energy of the flow is given by
\begin{equation}
\label{4.1}\mathcal{E}_f=\frac{1}{2}\rho\int_{V_f}\vc{u}^2\;d\vc{r},
\end{equation}
where $V_f$ indicates the part of space occupied by fluid.
The integral is absolutely convergent. The kinetic energy changes due to work done on the fluid by the moving body surface and by dissipation. The rate of change is given by
\begin{equation}
\label{4.2}\frac{d\mathcal{E}_f}{dt}=W-\mathcal{D},
\end{equation}
where $W$ is the work done by the body surface per unit time
\begin{equation}
\label{4.3}W=\int_S\vc{u}\cdot\vc{\Pi}\cdot\vc{n}\;dS,
\end{equation}
and $\mathcal{D}$ is the rate of dissipation given by
\begin{equation}
\label{4.4}\mathcal{D}=2\eta\int_{V_f}\vc{\Delta}:\vc{\Delta}\;d\vc{r},\qquad \vc{\Delta}=\frac{1}{2}((\nabla\vc{u})+(\nabla\vc{u})^T).
\end{equation}
Averaging Eq. (4.2) over a period we find that the mean power can be evaluated either from the work or from the dissipation, since
\begin{equation}
\label{4.5}\overline{W}=\overline{\mathcal{D}}.
\end{equation}
The calculation of the work is preferable, since it is given by a surface integral.
In the process of dissipation due to fluid viscosity, kinetic energy of flow is converted into internal energy. The internal energy created during a period is transported out of the system by heat conduction. This is left out of the description.

We note that if the flow is divided into a regular part and a wake, then the rate of change of the kinetic energy of the regular part is partly due to flow of energy from the regular part into the wake. Lighthill called this energy wastage \cite{30}. Eventually this energy is also converted into internal energy by dissipation.

The efficiency of swimming is defined by the ratio \cite{15}
 \begin{equation}
\label{4.6}E_T=4\eta\omega a^2\frac{\overline{U}}{\overline{\mathcal{D}}},
\end{equation}
where $a$ is a typical size of the swimmer.
For the swimmers specified in Eq. (2.11) we derived \cite{15},\cite{16} for the mean swimming velocity to second order in $\varepsilon$
\begin{equation}
\label{4.7}\overline{U^{(2)}}=\frac{1}{2}\varepsilon^2(l+1)(l+2)\mu_l\mu_{l+1}a\omega.
\end{equation}
and for the mean rate of dissipation
\begin{equation}
\label{4.8}\overline{\mathcal{D}^{(2)}}=4\pi\varepsilon^2[(l+1)(l+2)\mu_l^2+(l+2)(l+3)\mu_{l+1}^2]\eta\omega^2a^3.
\end{equation}
This yields for the efficiency
\begin{equation}
\label{4.9}E_T=\frac{1}{2\pi}\frac{(l+1)\mu_l\mu_{l+1}}{(l+1)\mu_l^2+(l+3)\mu_{l+1}^2}.
\end{equation}
The efficiency is maximal for $\mu_{l+1}=\sqrt{(l+1)/(l+3)}\mu_l$,
\begin{equation}
\label{4.10}E_{Tmax}=\frac{1}{4\pi}\sqrt{\frac{l+1}{l+3}},
\end{equation}
as found earlier \cite{15}.

Using the orthonormality relations for the vector spherical harmonics \cite{21},\cite{31} we find that to second order the kinetic energy of the flow depends on time as
\begin{equation}
\label{4.11}\mathcal{E}^{(2)}_f(t)=\varepsilon^2\pi\bigg[\frac{2l+2}{2l+1}\mu_l^2\cos^2\omega t+\frac{2l+4}{2l+3}\mu_{l+1}^2\sin^2\omega t\bigg]a^5\rho \omega^2.
\end{equation}
This is nearly constant in time.

These energy considerations pertain only to the fluid. The calculations can be performed from the flow, as given by the approximate solution of the Navier-Stokes equations. Again we leave the inner dynamics of the body out of consideration. In living organisms these are the concern of physiology. In particular, the total work and rate of dissipation will be relevant in the determination of the stroke of optimal efficiency \cite{32}.

\section{\label{V}Structure considerations}

We investigate briefly how the forces calculated above are related to the inner structure of the body. To second order the force exerted by the fluid on the body is given by Eqs. (2.14) and (3.9). By Newton's equation of motion,
  \begin{equation}
\label{5.1}\frac{d(m_0 \vc{U}_{cm})}{dt}=\vc{K}(t),
\end{equation}
where the total body mass $m_0$ usually is independent of time, this allows us to calculate the center of mass velocity $U_{cm}(t)=\overline{U}+\delta U_{cm}(t)$, where $\delta U_{cm}(t)$ is oscillating with period $T$. Added mass effects \cite{25},\cite{26} are embodied in the hydrodynamic force $\vc{K}(t)$, as in the equation of motion for a rigid sphere \cite{24}. The center of mass velocity $\delta U_{cm}(t)$ depends on the body mass and will differ from the swimming velocity $\delta U(t)=U(t)-\overline{U}$, where $U(t)$ is the mean velocity of the surface, which we calculated to second order on the basis of kinematics \cite{19}. Such a difference between the velocity of the hull and the center of mass velocity is employed in rowing \cite{27}-\cite{29} to optimize the mean velocity $\overline{U}$.

In the DQ model with constant mass $m_0$ Newton's equation reads to first order
  \begin{equation}
\label{5.2}m_0\frac{d\vc{U}^{(1)}_{cm}}{dt}=\vc{K}^{(1)}(t),
\end{equation}
with $\vc{K}^{(1)}(t)$ given by Eq. (2.14). On the other hand we have put $\vc{U}^{(1)}=0$, so that the hull is stationary to first order. We can imagine, for example, a mechanical model where the sphere is filled with a massless elastic material transmitting the force to a point mass $m_0$ bound elastically to the center of the sphere. The DQ model cannot be realized for a uniform swimmer. For the QO model the first order force vanishes, so that $\vc{U}^{(1)}_{cm}=0$.

For a spherical deformable body with some internal elastic structure and mass distribution immersed in the fluid we can in principle consider the linear response to an oscillating surface force density acting in proportion to the vector spherical harmonic $\vc{B}_1$ on each mass point at the surface. If the body were a simple elastic particle \cite{33}, \cite{34}  then the knowledge of the force distribution over the distorted surface would allow for a discussion of internal elastic modes. However the detailed internal structure of a living organism is much more complex \cite{35}, \cite{36} and is beyond the scope of the present study. We can say that the amplitude of displacement of the points at the surface is characterized by the complex vector
  \begin{equation}
\label{5.3}\vc{d}_\omega=a[R_S(\omega)\vc{e}_z+A(\omega)\vc{B}_1],
\end{equation}
with response factors $R_S(\omega),A(\omega)$, and corresponding velocity
  \begin{equation}
\label{5.4}\vc{U}_\omega=-ia\omega R_S(\omega)\vc{e}_z.
\end{equation}
The center of mass velocity will be
  \begin{equation}
\label{5.5}\vc{U}_{cm\omega}=-ia\omega R_{cm}(\omega)\vc{e}_z,
\end{equation}
with another response factor $R_{cm}(\omega)$. With no-slip boundary condition at the surface the flow velocity in the outer space is
  \begin{equation}
\label{5.6}\vc{u}_\omega(\vc{r})=\vc{v}_{U\omega}(\vc{r})-i\omega A(\omega)\frac{a^4}{r^3}\vc{B}_1(\theta),
\end{equation}
where $\vc{v}_{U\omega}(\vc{r})$ is the flow pattern for a rigid sphere moving with velocity $\vc{U}_\omega$, given explicitly by Eq. (4.3) of Ref. 19. Provided the response factor $R_S(\omega)$ is much less than $A(\omega)$ it can be neglected in Eq. (5.3). We recall that $\vc{B}_1=\vc{e}_z-3\cos\theta \vc{e}_r$. Correspondingly the first term in Eq. (5.6) can be neglected as well, and the first order outer flow is nearly dipolar.

\section{\label{VI}Thrust-drag paradigm}

It is worthwhile to study also the mean force density on the surface. According to Eq. (3.5) its integral over the surface vanishes. We shall consider separately the $z$ component of the pressure force density and of the viscous force density. It turns out that for both models the mean pressure force density is symmetric about polar angle $\theta=\pi/2$ and the mean viscous force density is antisymmetric about this angle. Again we do not consider the mean Reynolds force density and the corresponding contribution to the pressure, since these just cancel locally. The expressions below are calculated from the time average of the integrands in Eq. (2.16).

For the DQ model we find for the remaining pressure contribution to the mean force density at the surface $r=a$
  \begin{eqnarray}
\label{6.1}\overline{k_{pz}^{(2)}}&=&\frac{9}{8}\;\varepsilon^2\eta\omega\mu_1\mu_2\cos^2\theta(5\cos2\theta-1)\nonumber\\
&+&\frac{1}{64}\varepsilon^2a^2\rho\omega^2\cos\theta\big[-32\mu_1^2+21\mu_2^2+12(8\mu_1^2-5\mu_2^2)\cos2\theta+135\mu_2^2\cos4\theta\big].
\end{eqnarray}
The first term arises from the first integral in Eq. (2.16).
For the viscous contribution to the mean force density we find
  \begin{equation}
\label{6.2}\overline{k_{vz}^{(2)}}=\frac{9}{8}\varepsilon^2\eta\omega\mu_1\mu_2\sin^2\theta (3+5\cos2\theta).
\end{equation}
In Fig. 3 we plot the two force densities as functions of $\theta$ for $s=2$ and $\mu_1=1,\;\mu_2=1/\sqrt{2}$, corresponding to optimum efficiency, as shown above Eq. (4.10). The swimmer swims in the positive $z$ direction.
The expressions  yield for the force exerted on the front part of the sphere
  \begin{eqnarray}
\label{6.3}\overline{K^{(2)}_{pz1}}&=&2\pi a^2\int^{\pi/2}_0\overline{k_{pz}^{(2)}}\sin\theta\;d\theta=-\frac{\pi}{8}\varepsilon^2a^4\rho\omega^2[4\mu_1^2+3\mu_2^2],\nonumber\\
\overline{K^{(2)}_{vz1}}&=&2\pi a^2\int^{\pi/2}_0\overline{k_{vz}^{(2)}}\sin\theta\;d\theta=0.
\end{eqnarray}
The first term in Eq. (6.1) does not contribute.

For the QO model we find for the remaining pressure contribution to the mean force density at the surface $r=a$
\begin{eqnarray}
\label{6.4}\overline{k_{pz}^{(2)}}&=&\frac{5}{448}\;\varepsilon^2\eta\omega\mu_2\mu_3\cos^2\theta\big[-19+1148\cos2\theta+567\cos4\theta\big]\nonumber\\
&+&\frac{1}{64}\;\varepsilon^2a^2\rho\omega^2\cos\theta\big[21\mu_2^2-92\mu_3^2-15(4\mu_2^2-7\mu_3^2)\cos2\theta
+15(9\mu_2^2-4\mu_3^2)\cos4\theta+175\mu_3^2\cos6\theta\big].\nonumber\\
\end{eqnarray}
For the viscous contribution to the mean force density we find
\begin{equation}
\label{6.5}\overline{k_{vz}^{(2)}}=\frac{5}{448}\;\varepsilon^2\eta\omega\mu_2\mu_3\sin^2\theta\big[1245+2156\cos2\theta+567\cos4\theta\big].
\end{equation}
In Fig. 4 we plot the two force densities as functions of $\theta$ for $s=2$ and $\mu_2=1,\;\mu_3=\sqrt{3/5}$, corresponding to optimum efficiency, as shown above Eq. (4.10).
The expressions  yield for the force exerted on the front part of the sphere
  \begin{eqnarray}
\label{6.6}\overline{K^{(2)}_{pz1}}&=&2\pi a^2\int^{\pi/2}_0\overline{k_{pz}^{(2)}}\sin\theta\;d\theta=-\frac{3\pi}{8}\varepsilon^2a^4\rho\omega^2[\mu_2^2+3\mu_3^2],\nonumber\\
\overline{K^{(2)}_{vz1}}&=&2\pi a^2\int^{\pi/2}_0\overline{k_{vz}^{(2)}}\sin\theta\;d\theta=0.
\end{eqnarray}
The first term in Eq. (6.4) does not contribute.

For both models the viscous force $\overline{K^{(2)}_{vz1}}$ vanishes. This must be so on account of our earlier results in Eqs. (3.1) and (3.4), and the symmetry about $\theta=\pi/2$. In both cases the mean force acting at the back part of the surface is given by
\begin{equation}
\label{6.7}\overline{K^{(2)}_{pz2}}=-\overline{K^{(2)}_{pz1}}.
\end{equation}
This is in the direction of swimming. The equality implies that the mean propulsive force acting on the back of the surface is balanced by an opposite braking force on the front.

The models show that the thrust-drag paradigm should be viewed with skepticism. Viscous drag does play a role in the models, implying that momentum is absorbed by friction, but the net viscous force, i.e. the total drag, vanishes. The total mean pressure force vanishes also.

Unlike in towing of a rigid body, in swimming there is no net transfer of momentum to infinity. This implies that the mean total force vanishes. One may think of swimming as a process of momentum exchange between body and fluid resulting in a steady net forward motion. It is useful to divide the body surface into two parts, one predominantly creating fluid momentum in the backward direction, and the other part predominantly absorbing this again.

The thrust in swimming may be identified with the mean force corresponding to the momentum generating part. The definition of thrust is subjective, since the dividing line between the two parts can be chosen judiciously. For our two models the natural choice is the equator on the sphere. In human swimming the momentum generating part might be considered to be the inside of the hands and the bottom of the feet, the momentum absorbing part being the rest of the body. In swimming involving a wagging tail or a flapping wing it makes sense to switch the dividing line at some time during the period. Depending on the stroke the dividing line may consist of several parts, and switching may be done for some of the parts.

The thrust may be due to viscous friction and/or to pressure forces, and similarly for the momentum absorption on the remaining part of the surface. For our two models only the pressure force contributes to the mean thrust and to the mean momentum absorption. Other swimmers may rely partly or entirely on viscous forces for the generation of thrust \cite{37}. In the models there is no relation between the calculated expression for the force $\overline{K^{(2)}_{pz2}}$ and the mean swimming velocity given by Eq. (4.7). The latter was found from a kinematic calculation employing the condition that the total mean force vanishes. The models suggest that in laminar swimming the calculation of thrust is irrelevant for the mean swimming velocity. Conceivably the situation is different at high Reynolds number, with transition to a regime where the total mean force does not vanish.

\section{\label{VII}Discussion}

In earlier work we studied cruising swimming at small amplitude as a boundary value problem in hydrodynamics \cite{15}. In principle the periodic motion of the fluid can be determined from the Navier-Stokes equations for prescribed distortions of the deformable boundary. For the DQ and QO models this can be demonstrated in detail. In particular the mean swimming velocity and the mean rate of dissipation can be evaluated explicitly. In the above we showed that also the time-dependent force exerted by the fluid on the body can be found. This allows calculation of the time-dependent center of mass velocity. For the two models this differs from the time-dependent velocity of the body surface, though of course the time-average is the same.

The analysis leads to an important conclusion. If the surface distortion has a dipolar component, then a time-dependent force, linear in the amplitude of the stroke, is exerted by the fluid on the body. This leads to an oscillating first order swimming velocity, unless the dynamic structure of the body is such that the force is fully absorbed internally, without effect on the surface motion, as discussed in Sec. V. In particular for a uniform body the surface velocity and the center of mass velocity are the same. The first order velocity for a uniform distorting sphere is discussed elsewhere \cite{38}.

In simplified mechanical models \cite{9}-\cite{12} the full dynamics of the body is taken into account, but at the expense of a simplified description of hydrodynamic interactions. The DQ and QO models have the advantage that the dynamics of the fluid is treated exactly, though only to second order in the amplitude of the stroke and with the neglect of turbulence. It would clearly be of interest to extend the calculation to the fully nonlinear regime.

\newpage

\newpage

\section*{Figure captions}

\subsection*{Fig. 1}
Plot of the ratio $K^{(2)}_p(t)/K_{01}$ (short dashes), $K^{(2)}_v(t)/K_{01}$ (long dashes), as well as their sum (solid curve) for scale number $s=0.01$ (left panel), $s=1$ (middle panel), and $s=100$ (right panel) as functions of $\omega t$ for the DQ model.

\subsection*{Fig. 2}
Plot of the ratio $K^{(2)}_p(t)/K_{02}$ (short dashes), $K^{(2)}_v(t)/K_{02}$ (long dashes), as well as their sum (solid curve) for scale number $s=0.01$ (left panel), $s=1$ (middle panel), and $s=100$ (right panel) as functions of $\omega t$ for the QO model.

\subsection*{Fig. 3}
Plot of the reduced force density $\overline{k_{pz}^{(2)}}/(\varepsilon^2\eta\omega\mu_1\mu_2)$ for $s=2$ (solid curve) and of $\overline{k_{vz}^{(2)}}/(\varepsilon^2\eta\omega\mu_1\mu_2)$ (dashed curve) for the DQ swimmer with $\mu_2=\mu_1/\sqrt{2}$.

\subsection*{Fig. 4}
Plot of the reduced force density $\overline{k_{pz}^{(2)}}/(\varepsilon^2\eta\omega\mu_2\mu_3)$ for $s=2$ (solid curve) and of $\overline{k_{vz}^{(2)}}/(\varepsilon^2\eta\omega\mu_2\mu_3)$ (dashed curve) for the QO swimmer with $\mu_3=\mu_2\sqrt{3/5}$.

\newpage
\setlength{\unitlength}{1cm}
\begin{figure}
 \includegraphics{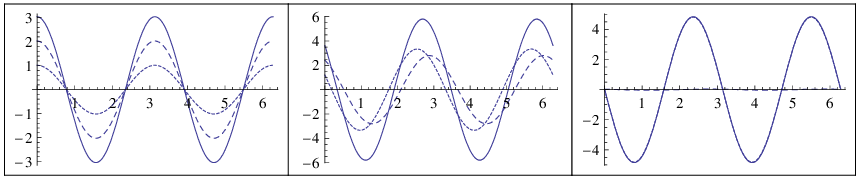}
   \put(-9.1,3.1){}
\put(-1.2,-.2){}
  \caption{}
\end{figure}
\newpage
\clearpage
\newpage
\setlength{\unitlength}{1cm}
\begin{figure}
 \includegraphics{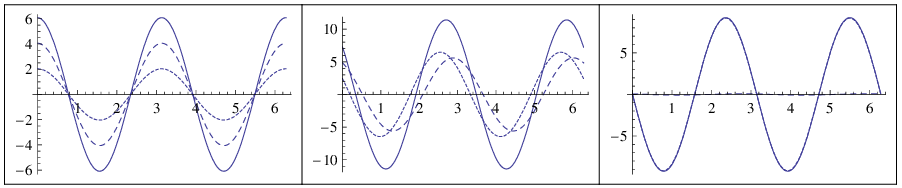}
   \put(-9.1,3.1){}
\put(-1.2,-.2){}
  \caption{}
\end{figure}
\newpage
\clearpage
\newpage
\setlength{\unitlength}{1cm}
\begin{figure}
 \includegraphics{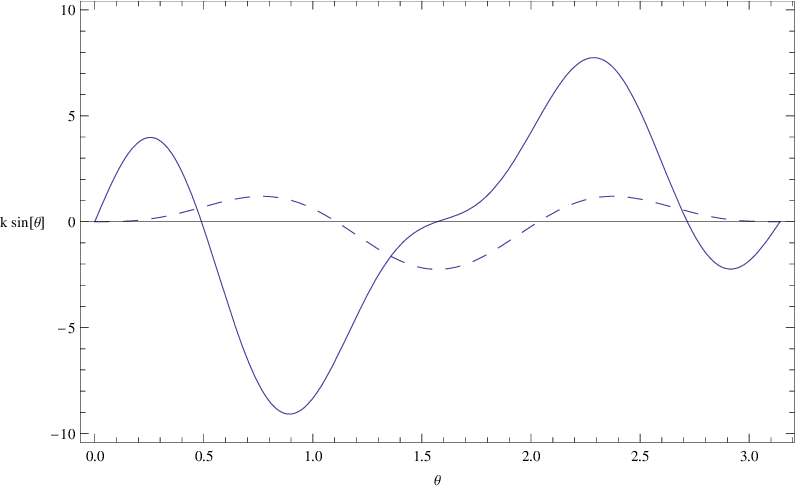}
   \put(-9.1,3.1){}
\put(-1.2,-.2){}
  \caption{}
\end{figure}
\newpage
\clearpage
\newpage
\setlength{\unitlength}{1cm}
\begin{figure}
 \includegraphics{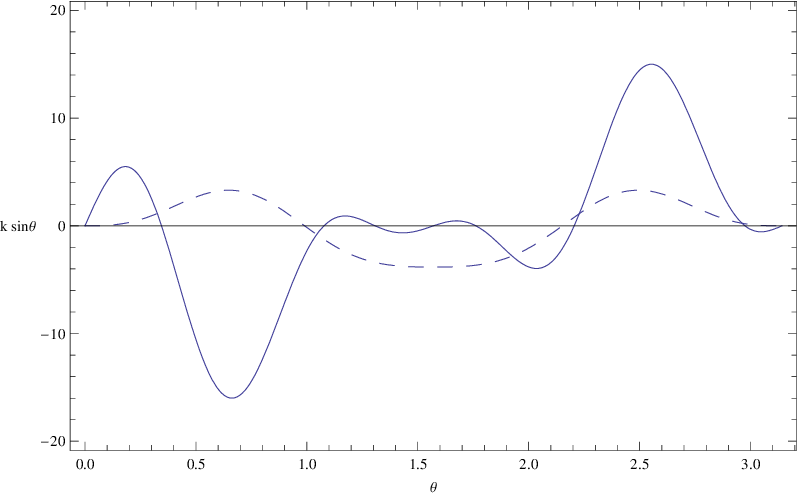}
   \put(-9.1,3.1){}
\put(-1.2,-.2){}
  \caption{}
\end{figure}
\newpage
\end{document}